# REJOINDER OF: STATISTICAL ANALYSIS OF AN ARCHEOLOGICAL FIND

By Andrey Feuerverger

*University of Toronto*

I thank all the discussants for their many critiques and comments, and for their considerable efforts. Many of the points raised are ones with which I (at least in part) agree. It therefore seems easiest to first deal with a number of points with which I don't agree.

First, Fuchs states (and Bentley appears to assume) that my analysis is documented in a book and in a movie, neither of which I have authored. In fact, it is documented *only* in my paper which references neither of these, and neither does it reference any developments which occurred subsequent to my work. Although I will need to comment on one such development below, I otherwise confine this reply to the contents of my paper and to those comments of the discussants which appear within this issue of the *Annals*. In particular, I avoid being drawn here into discussions concerning representations made elsewhere by others, or to any matters alluded to by discussants that are peripheral to the central and substantive statistical issues of the problem. Nothing in this work was ever intended to cause offence to anyone. In my view, the statistical problems here are of methodological interest, and the subject matter is one of historical and archeological significance. If this tomb is not that of the NT family (as indeed it may not be) then archaeological work could still one day unearth a tomb that is and the question of what statistics might then contribute toward such a pursuit could then become important.

I also want to say that my paper does not—as some discussants intimate— claim that the Talpiyot tomb "is most likely that of the NT family." What it tries to do is develop tools to assist subject matter experts in their work of gauging the veracity of any such claims. The function of statistics here is to help out in the difficult historical and archeological work. The critical role which historical assumptions play here means that such calls are not ours to make; and like Fuchs, I too refrain from passing judgment on the subject matter issue of whether this is or is not the NT tombsite. Of course, after the fact, it is easy to gain a sharpened appreciation for the safety of a "nihilistic" approach, one that—as Höfling and Wasserman put it—provides no answers. However, the intellectual temptations posed by a problem of this nature are surely too great to simply set aside.

*Meaning of surprise.* Turning to some specific issues raised by the discussants, I think it is important to distinguish more carefully between "interesting" or "relevant" collections of names, and what I have defined as being "surprising" collections of names. If a NT tombsite actually exists, it is certainly within the realm of prior possibilities that it contains within it only the most common renditions for the names of persons who might be recognizable to us. Were this so, no purely statistical procedure would then be able to "detect" it because such collections of names would not occur rarely enough in the general population to allow any procedure at least an opportunity to attain significance—that is, we then could never know. Indeed, only if the actual burials had taken place under rarer relevant renditions of the names (and only if in a tomb of a certain size) could there ever be a chance to "detect" it statistically. In other words, some historical and archeological "good fortune" would also be required.

Both Höfling and Wasserman, as well as Fuchs, appear to misinterpret my definition of "surprisingness" and its intent. In fact Höfling and Wasserman state that "the RR statistic becomes more significant if broad name categories are being subdivided into special name renditions, even if the particular name renditions are not relevant." But that does not take into account that the specialness of a name rendition is permitted to count only if it is relevant, and only if it appears in a prespecified nested list of increasingly more specialized (i.e., "rarer and more relevant") name renditions. The rareness alone of a name rendition (even if it corresponds to a generic name category deemed to be highly relevant) is not of essence. When Höfling and Wasserman state that "interested observers would surely argue that a tomb is interesting if there is any way at all of matching the names found to potentially interesting names," they bypass the fact that such matchings will be relatively too probable to be significant if they were to occur under the most common renditions for the names. Likewise, Fuchs suggests in an example that, had the Talpiyot tomb contained a Salome in lieu of the Mariamenou [$\eta$] Mara inscription, it would have been considered still more surprising even though its RR value would then have been higher. In fact, based on the definition of surprisingness, had a Salome been found in lieu of the Mariamenou inscription, the cluster might conceivably be described as being more relevant, but (in view of how common Salome was as a name) it certainly would not have been more surprising, that is, it would not have provided a greater evidentiary value (under our provisos).

*Other misinterpretations.* Fuchs also remarks that if a Simon had been found in the tomb instead of the Matya, our RR value, and hence our tail areas, would have been unchanged even though that Simon might have been a brother of Jesus. Now one can certainly carry out analyses allowing for the brother Simon to be a candidate for a NT tomb. The reason we did not do so, however, is that Simon is presumed to have died subsequent to the time (70 CE) when the practice of ossuary burial ceased. If this is so then it is appropriate to have, as Fuchs puts it, "ignored that inscription."

Fuchs also states that assigning an RR value of 1 to names in the "Other" category means that such names will not contribute to the RR value of a cluster, the implication being that such names are then simply ignored by our procedure. This actually is not so. Such values of 1 do in fact contribute in the sense of being values much higher than would have occurred had relevant names been encountered instead. (As well, such values of 1 for "Other" also affect our null distribution.) And in any case, we have also allowed for values exceeding 1 via the device of a disqualifier list, but we did not implement such a list since no such names appeared in-sample; disregarding such names was therefore conservative.

One referee of the paper had stated that merely including the Talpiyot names within the onomasticon necessarily biases our results. I do not agree that this is the case. Any and all available names may and should in fact be used to aid in the process of constructing the prespecified categories of name renditions and nesting them according to "relevance and rareness." What is important is that this process be carried out without reference to which of the names actually originated from within the tomb.

Fuchs mentions that the RR value for the Talpiyot tomb uses only four out of the six inscribed ossuaries on account of the fact that the Matya ossuary only contributes a 1 to the RR value for the tomb. That view is not entirely correct. As noted above, the Matya ossuary renders the RR value for the tomb much higher than it would have been had a more relevant name been encountered in its place, and in turn this results in a considerably increased tail area (for the RR value) under the null distribution. The situation with respect to the sixth inscription (i.e., Yehuda bar Yeshua) will be taken up separately below; for the moment we only note that, in effect, this inscription also contributes an RR value of 1.

*A hypothesis testing issue.* Fuchs states that "A set of rules which weigh positively (i.e., with a coefficient less than 1) names expected under $H_1$, but does not weigh negatively names which are unexpected under $H_1$, is likely to yield biased results in favor of $H_1$" and that "this procedure is at least questionable." Kadane's critique of the RR measure points in a related direction. Intuitive as such remarks may seem, however, they are not entirely correct. First, as long as a test statistic is specified a priori (in particular, without making reference to the data), and as long as its distribution under the null hypothesis is specified correctly, the resulting test will be unbiased in the sense of having its stated level of significance. Essentially, only the power of the test will be at issue—a consideration that leads us to seek procedures with high ability to discriminate. Second, our allowance for a disqualification set does to some extent permit certain names to weigh negatively; the reason we did not implement such a set (as mentioned previously) was only because Matya was not considered to be a disqualifying name so that our choice not to do so was conservative. (In fact, the Matya name was viewed to be neutral.) Contrary to Fuchs' assertions, there is nothing questionable about our actual *procedure*. The matter of the Yehuda inscription will be discussed separately below.

*Post hoc inference.* Fuchs does point out correctly that "the a priori nature of the provisos is amongst the most important premises" of the analysis. He further states that "The overall impression is that the inevitable exposure to the data affected the definition of the provisos." Concerning the degree to which the provisos were truly a priori, he adds: "It is difficult to accept that ... the elements ... have indeed been so specified." On such points I am certainly sympathetic to the general nature of the concerns raised by the discussants and therefore revisit certain of the 'provisos' in the discussions below. It is indeed true here that prior exposure to the data was inevitable, and the principle is well understood that biases result if data is used when setting up an inference. In fact, the best we have been able to do was to stress the fact that the data has been seen. And we also tried, both conscientiously and hard, and on a best efforts basis, to construct our inference to be as a priori as possible in the circumstances. The extent to which we have succeeded or not in this task is one which each reader ultimately must judge for themselves. For regardless of the degree of objectivity any analyst may wish or seek to claim in carrying out an analysis under such circumstances, no convincing or irrefutable proof of such objectivity can ever be offered. This is the perennial problem of pre- versus post-hoc inference, and the present statistical problem provides a good example of it. It is also the reason why I tried to be so careful to isolate and exhibit all of the assumptions under which the analysis was carried out.

*Mariamenou need not be Mary Magdalene.* Before addressing the critiques to my provisos, there is one further item that needs to be clarified. Fuchs intimates that the RR value assigned to the Mariamenou [$\eta$] Mara inscription means that this name must then necessarily refer to Mary Magdalene. (A related remark is also made by Bentley.) This interpretation is not correct. That RR value resulted from that version of the name having been considered, on an a priori basis, to be the most 'rare and relevant' rendition of the her name from amongst those names that we know. (But on this point, note the further discussion below.) It assumes no more than that only one woman having the generic name category of Mariam out of about every 44 such women could legitimately have been called by that rendition, and that Mary Magdalene was among those who could so be called. In particular—although that possibility was weighed into the process when deciding upon our a priori nested rendition categories—it certainly was *not* assumed that the Mara in the inscription *must* be an honorary title, only that it *might* be. For if that were assumed one must surely agree with Fuchs (and Bentley) that no statistical analysis would then be required.

We now turn our focus to some of our various provisos starting with the ones associated with Mary Magdalene.

*Mary Magdalene as a priori candidate.* With respect to Mary Magdalene, there are at least two distinct considerations. The first of these is the matter of her inclusion on our list of a priori candidates for a NT tomb. Of course, this is primarily a historical issue, not a statistical one, and as such needs to be vetted through *dispassionate* subject matter expertise. While sensitivities surrounding this point render the scholarly work more difficult, I really do not see how one can exclude her from that list. This is in no way tantamount to any assumptions about to whom, if anyone, she may have been married. The perceptions of Mary Magdalene having been unchaste apparently originates with Pope Gregory the Great in the last decade of the sixth century and has no basis at all in the NT—a point that even the Vatican conceded in 1969. Her presence is felt both prominently and strategically throughout the NT accounts. She is the pivotal figure and primary source for the resurrection. She accompanies Jesus over substantial distances and over a substantial period of time. She even appears, from the accounts, to have been highly active in Jesus' ministry. She is present at the crucifixion, and also at the burial where (in view of the likely nature of such rituals in that era) one would expect only intimates of the family to attend. Indeed, she is also cited as having been in the vicinity of the tombsite on multiple occasions. So, on balance—and in view of the possibility that she may have been buried in the Holy Land—I really do not see how one can realistically exclude her from at least being a *candidate* for a NT tombsite. One must avoid a certain blurring of logic that can occur inadvertently here: The inclusion of Mary Magdalene on an a priori list of candidates for a NT tombsite is *not* equivalent to asserting that she must actually be found in such a tomb. It only says that she is, a priori, among the plausible *candidates.* The distinction between these must not be blurred by the occurrence of the Mariamenou [η] Mara ossuary within our data. Of course, there is no obligation on anyone's part to accept the argument we make here; if one chooses to omit Mary Magdalene as an a priori candidate then the impact of that choice is clear: no statistical analysis applied to the Talpiyot data would then attain significance.

*Names for Mary Magdalene.* With respect to the second key consideration pertaining to Mary Magdalene, the situation is more problematic. At the time I did the analysis my "due diligence" in respect of constructing an a priori nested list of name renditions for Mary Magdalene included such elements as the following:

(a) The itemization of the 80 known renditions for the generic name of Mariam as recorded in Ilan (2002).

(b) The meaning of the Aramaic word "mara"; specifically, Ilan (pp. 392 and 423) states: "Mara means 'lord, master' in Aramaic."

(c) An inference, based on (a) and (b), that of these 80 name renditions for Mariam, the extraordinary Mariamenou [$\eta$] Mara one may provide an arguably tighter "fit" to Mary Magdalene than any of the other 79.

(d) The reference by Hippolytus, around 175 CE, in his Refutations, to a particular Mariamne who was legitimately a "mara" in the Aramaic sense of that word.

(e) The unwavering opinion of Rahmani, and of some other highly regarded epigraphers (e.g., Leah Di Segni), that the full inscription on ossuary #1 was intended to refer to a single individual only.

(f) The article by Francois Bovon (2002) identifying the Mariamne in the Acts of Philip as Mary Magdalene, and identifying Philip as her brother.

And finally:

(g) Information provided to me (but note the discussions below) that Professor Bovon—a highly respected scholar and expert on this subject—was on record as having authenticated that Mariamne was most likely the actual name of Mary Magdalene.

I was, of course, also aware of the fact that the inscription Mariamenou [$\eta$] Mara had occurred within the tomb, and obviously also of the fact that such information must be disregarded when forming a priori assumptions. However there is as well a concomitant piece of information of a seemingly ancillary kind, and not entirely unrelated to our conditioning on the tomb's configuration. Namely, it is known that the Mariamenou [$\eta$] Mara inscription was rendered in Greek, but that it occurred within a tombsite containing five other inscribed ossuaries all of which were rendered in Aramaic. How and if such a piece of information may be used in forming a priori assumptions is not entirely clear to me and I leave this as a question for readers to consider. Similar issues also arise in respect of such considerations as the nature of the actual incisions and so on. I also point out in passing—although this should not be regarded as being an a priori observation—that to the best of my knowledge, Mary Magdalene is the only historical personage who was ever referred to by the generic name of Mariam combined with the Hebrew letter "nun," and that she is referred to in that way in two distinct sources (Hippolytus and the Acts of Philip).

The controversies resulting from the airing of the documentary film was a unique event in the context of any statistical problem I had ever dealt with, and went beyond what I might realistically have been able to prepare for. Scholars and others who were involved in any way were subjected to pressures that sometimes made it difficult to discern where the actual facts lay. Speaking for myself, I was interested only in the facts. The story of the crucifixion has held sway over the history of humanity for some 2000 years. It therefore seemed worthwhile to stay the course that happenstance had led me to, and to steadfastly pursue the facts to whatever would be their logical conclusion.

*Bovon's clarification.* This brings us to the subject of the clarifications subsequently issued by Professor Bovon. There is no doubt whatever now that these were not retractions in response to pressures nor were they motivated by a recognition of the possible uses which might be made of such work. In fact, Bovon's clarifications are those of a serious scholar whose remarks—having inadvertently been misinterpreted by Jacobovici—were conveyed to me out of context. To quote from Bovon's statement to the Society of Biblical Literature:

"I do not believe that Mariamne is the real name of Mary of Magdalene. Mariamne is, besides Maria or Mariam, a possible Greek equivalent, attested by Josephus, Origen, and the *Acts of Philip*, for the Semitic Myriam."

"Mariamne of the *Acts of Philip* is part of the apostolic team with Philip and Bartholomew; she teaches and baptizes. In the beginning, her faith is stronger than Philip's faith. This portrayal of Mariamne fits very well with the portrayal of Mary of Magdala in the Manichean Psalms, the Gospel of Mary, and Pistis Sophia. My interest is not historical, but on the level of literary traditions."

Without benefit of the last element, that is, (g), of the itemization above, I do not regard the assumption A.7—concerning the most appropriate name rendition for Mary Magdalene—as being equally adequately justified by the remaining elements (a) through (f) on that list. In particular, this means that we cannot (on the basis of our RR procedure) say that the Talpiyot find is statistically significant in any meaningful way. Readers who wish to form their own judgement on this should note that the germane question here is not whether or not Mariamne was the actual name of Mary Magdalene, but whether or not we are justified—on an a priori basis—to say that the rendition Mariamenou [$\eta$] Mara provides a better fit to the name of Mary Magdalene than any of the others, whilst bearing in mind that she is repeatedly referred to in the NT as having come from Migdal, and is not referred to there as Mariamne. We shall see below, however, that this matter is not yet closed.

*The Yehuda ossuary.* Now let us deal with the matter of the sixth ossuary—the admittedly problematical one inscribed Yehuda bar Yeshua. When I encountered this data set I did not at first have a clear idea of how that datum should be dealt with in an analysis and I tentatively set it aside. It would be fair to say that the apparent implications suggested by that ossuary would hardly have found any mention of or allowance for in my list of a priori assumptions for several reasons, not the least of which being that such a possibility would not ever have occurred to me. After the RR approach evolved, it became clear to me that this sixth ossuary was actually being incorporated within the computations in a particular way. As indicated in Section 14 of the paper, the analysis may in fact be carried out allowing for the presence of a generationally aligned sequence of the form "A son of B son of C" with the youngest of this trio "not counting" toward the RR value due to our lack of knowledge about *any* father-and-son pair *both* dying within the 30–70 CE timeframe. Of course, this still leaves open the question of associated a priori assumptions. If one ascribes to certain theological interpretations later placed upon the historical events, the decision is clear: the outcome observed must belong to the disqualification set, and the matter is closed. If one does not so ascribe, the situation becomes more difficult, for then one must interpret the historical records as best one can to assess the plausibility of such an outcome, and address such questions as the following: Would a union in such an instance have been sanctioned? Was it— in that era—viewed as improper to father a son? Did Jesus advocate against it for either self or followers? If there were a son, would there have been a recognized threat to his life? We cannot answer these or other such questions on behalf of the reader. Certainly the NT does not record any union or any son (although much other information is left unprovided as well). As for the statistical analysis based on RR, what we can say is that in assigning an RR value of 1 to the sixth ossuary, our procedure in effect acts with absolute neutrality on this question.

*Some extensions.* A few methodological points seem worth noting. Instead of prespecifying nested collections of name renditions one can (for each candidate individual) preassign numerical RR or surprisingness ratings to each of the onomasticon entries under their generic name. Only comparative (not numerical) values would actually matter, and the RR computation for an encountered rendition would then be based on the "tail area" resulting within the generic name category. Since many entries in the generic collection will have identical ratings, the resulting "discreteness" of the tail areas would act much as in our nested collections approach but would allow for somewhat greater flexibility. Note also that we could allow for the existence of rare renditions as yet unknown. For any candidate individual, the rareness of any such renditions would at most be (in the order of) that of a single unique entry in the onomasticon. We should remark here that a certain amount of variability in our results is attributable to the fact that name proportions are derived from the onomasticon which itself constitutes only a sample; Mortera and Vicard propose one method for assessing such variabilities.

*Other explanations and concerns.* Although many of the discussants focus on critiques to the analysis that might have been anticipated to arise out of theological grounds, Stigler alludes to some which stem from nontheological sources. For example—although the circumstances of the find assure us that the tomb had been undisturbed for many centuries—we do know that the Talpiyot tomb had been accessed at some point in antiquity. While it seems implausible to assume undue efforts on the part of those who did so, suppose they had found there only five of our six inscribed ossuaries and "recognized" the names on them. Might they not have thought it amusing to then take one of the uninscribed ossuaries there and crudely scratch upon it the name Yeshua bar Yehosef using an implement at hand? As for Stigler's reference to Sherlock Holmes' dog who did not bark, I did independently pursue the matter of why the placement of the ossuaries among the *kochim* had not been noted and concluded that this likely had occurred only on account of the general circumstances of the find and of Gath's untimely death upon which that potentially priceless piece of information was permanently lost. It must be remembered that the archaeologists who were sent there were not statisticians, that they could hardly have anticipated the nature of the questions that would later arise from this duty, that they had limited time inside a tomb containing only seemingly typical names, and that the messy Yeshua inscription could hardly have been decipherable to them at first. In fact one could (following up on a comment made by Bird) argue equally (although for what I believe are good reasons I do not) that the lengthy period which elapsed between the time of the tomb's discovery, and the time of the publication of its details, provides a yet contrasting instance of the dog not barking.

Stigler also raises the matter of our specialized independence assumption A.9. Our concerns, as well as our reasoning about this assumption, were discussed in Sections 7 and 14 of the paper. But in bringing this data set to the attention of the statistical community, it was understood that questions which merit further study would arise from it and the issue of cross-sectional independence is one of them. Here the question is not whether or not this assumption is true; we *know* that it is not. The question is whether the nature of that dependence affects the null distribution in an essential and nonconservative manner. I refrain from any rejoinders to Stigler's references to Bruno and Galileo finding such remarks too frightening to even contemplate.

Explanations based on coincidence should also not be overlooked; indeed, perhaps these data can be assessed under the framework of Diaconis and Mosteller (1989). Within the context of coincidence, odds of 1000 to 1 are hardly uncommon. Three "coincidences" weighed substantively in our analysis. One is the ossuary of Yeshua bar Yehosef. Another is the match to the rare name version Yoseh. And the third is the remarkable Mariamenou [$\eta$] Mara inscription. There are, however, also three further coincidences that (for reasons stated in the paper) I did not incorporate in the analysis but nevertheless seem worth noting. The first of these is the generational alignment of the three names Yehosef, Yeshua and Yehuda, with the alignment at Talpiyot being the only one among the six not immediately inconsistent with the NT family. The second is the seemingly suggestive choice, among the six ossuaries, on which the Greek script actually occurred, with the other five having been in Aramaic. And the third is the suggestive choice for which of the six ossuaries bore the messiest inscription—that choice being seemingly consistent with some theories that might be advanced to account for the empty tomb. Finally, there is yet one further coincidence: The youngest member of the generationally aligned ossuaries—namely Yehuda—has the same name as the youngest (or second youngest) brother of Jesus, with the accounts of Mark and Matthew having curiously reversed their two names.

Let us now address some specific further matters raised by the discussants.

*Höfling/Wasserman's first method.* In the "Different Approach" proposed by Höfling and Wasserman, the most essential difference actually lies in the treatment it accords to the different name versions. In particular they "lump together different versions of names" arguing "that a tomb is interesting if there is any way at all of matching the found names to potentially interesting names." Unfortunately, for common names, "interesting" will not be enough; there will be little opportunity for detection (i.e., the power will be low throughout all of the alternative) unless the renditions which occur match more specifically to the NT individuals, and if the specificness of such renditions is appropriately accounted for. A manifestation of this is that their calculation is "invariant under splitting names into subcategories," while our calculation (which attempts to account for the degree of rareness and relevance among the possible renditions) is not. Thus, had an inscription such as (say) "Yeshua of Nazareth, son of Yehosef" occurred in the tomb, their computations would be indifferent to an essential aspect of the name. Incidentally, Höfling and Wasserman are not correct in suggesting that what I have computed is "the probability of getting this set of names."

*Bayesian notions.* Several referees argue in favor of a Bayesian approach, something I tried to avoid due to the great divergences expected amongst priors (some of which have been influenced by theological considerations). Also, I do not entirely understand Kadane's remark about violation of the likelihood principle. Kadane appears to suggest that the uncertainties in deciding between whether a null hypothesis is false or whether a rare event has been observed is merely an artifact of the frequentist approach. It seems to me, however, that no purely statistical method can ever circumvent its analogue for "type 1 error." Further, in allowing a prior to place a zero probability on a discrete event, Kadane highlights a difficulty that can arise in a purely classical Bayesian approach, unless one takes to its extreme the view that "coherence" alone must suffice. It is also not entirely clear to me how straightforward it would be to implement LR procedures of the type Mortera and Vicard advocate. The Bayesian approach proposed by Höfling and Wasserman, however, does on first glance appear to lead to results comparable to those of a frequentist approach, as long as the assumptions under which the two approaches are implemented—in particular the assumptions concerning the renditions for the relevant names—are taken to be similar. Bayesian-like ideas may of course also be used to rationally combine subjective beliefs about individual assumptions into a plausibility for the collection of all assumptions. Our approach has been, however, for the RR method to act as a measuring device, to be tuned by the investigator in accordance with his or her expert assumption set.

*Bentley.* Bentley is correct in stating that my analysis assumes that a NT tomb might exist, but I do not fully agree with him that my analysis is conditioned on the assumption that such a tomb *must* exist with probability one. Also, while it may be fair of Bentley to argue with the estimates I used for the number of tombsites in Jerusalem, I am not aware of any expert opinion suggesting that the true number of tombsites is greatly in excess of the numbers I had used. Bentley's critiques regarding the Mariamenou [$\eta$] Mara inscription are well taken and this matter has been dealt with at length in our discussions above. If, in spite of my labors, Bentley wishes to be critical of them, he is within his rights. Nonetheless —lest Bentley's comments regarding James Tabor be misconstrued—I wish to say that in my discussions with Professor Tabor I found him always to be a scholar of impeccable integrity. Some of Bentley's comments, for example his closing remarks about archeologists and archeology being now at odds with statistics and with statisticians admittedly make for provocative and dramatic reading; unfortunately pressures of time do not permit me to enter into such debates.

*Ingermanson.* Ingermanson energetically presents "the case against" for essentially each one of the assumptions under which my computations were carried out. Although his critiques seem occasionally overzealous to me, they do provide a useful checklist of items that should be considered by anyone who seeks to arrive at a fully informed opinion about the Talpiyot tomb. Needless to say, an analysis of such data needs to be carried out under assumptions that are reasonable and defensible, even though no single assumption can ever be regarded as absolutely unassailable. Having already discussed many of such matters in my paper and throughout this reply, I do not repeat those arguments here, but instead return to two specific items. The first item concerns the treatment of the name Yoseh mentioned by Ingermanson (as well as others). I add here two additional points to those already made in the paper. First, surnames were not typical in that era and exceedingly common names (such as Yehosef) would not have provided adequate differentiation amongst individuals. In that respect, one needs to bear in mind the distinction between what we commonly refer to as being a "nickname" versus a name rendition or variant that is in itself intended to act as an actual name. (An instance of this may, e.g., have occurred in the case of the NT family.) Second, there is no singleton Yehosef inscription occurring within this tomb. Therefore such a seemingly more "formal" or more "respectful" version for Yoseh would have been available for use by the family without any risks of confusion but they chose not to use it. The second item concerns the name of the mother. I take issue with the objections Ingermanson (and others) raise regarding the a priori "rarer and more relevant" rendition of her name used in my analysis. The very earliest historical reference to her appears in Mark 6:3: "Is this not the carpenter/builder, the son of Maria, the brother of..." The second earliest reference also occurs in Mark when he mentions Maria as being at the cross, although whether or not this Maria ("the mother of James and Joseph") is meant to be Jesus' mother is not entirely certain. The third earliest historical reference to her appears in Matthew 1:18: "...when His mother Maria had been betrothed to Joseph..." (Luke does not use the form Maria but rather Mariam, however Luke is historically a significantly later source.)

*DNA and other evidence.* Mortera and Vicard raise the question of why DNA evidence was not collected and assessed more broadly and indicate some possible uses of such data. Bird alludes to some related matters as well. Having had reservations, such as about the risks of contamination, these were evidentiary points I decided not to pursue. I understand that it is, in any event, the case that such data cannot actually be obtained. While I appreciate the reasons behind such concerns, I also believe that Bird may be making more of the missing tenth ossuary than may actually be warranted by the facts.

*Critiques.* The critiques of the discussants encompass both the methodology and the assumptions under which it is being applied. The vigor of their remarks represents an important component of the scientific process when results seen to be controversial are being assessed. Although some of the discussants hold strong prior views on the subject matter, all critiques do nevertheless need to be considered on their own merits. So far as the methodology itself is concerned, I think I have addressed the main points that have been raised; however, the situation regarding the assumptions is necessarily different. These need to be vetted by *dispassionate* subject matter expertise. It is a curious and perhaps unique feature of this problem, however, that the body of subject matter expertise here is itself divided along very particular lines.

*A symposium.* In January 2008, at about the time I prepared this reply to discussion, I had the privilege to attend The Third Princeton Symposium on Judaism and Christian Origins held in Jerusalem. Several of the sessions at this conference were connected to matters relevant to evaluating the context of the Talpiyot tomb. Among the subject matter participants there working primarily with historical approaches, some indicated that they did not regard the Talpiyot tomb's being that of the NT family as an impossibility. All of the participants however (myself included) indicated that they did not regard that possibility as having been proven. The most interesting session there—relative to the requirements of our statistical analysis—was one on the epigraphy of the Talpiyot ossuaries during which the Mariamenou [$\eta$] Mara inscription was discussed. As might be expected, no consensus was reached in that session, but one remarkable *possibility* emerged of which few members in the audience (which consisted of nonstatisticians) grasped the immediate significance. That possibility, raised by Jonathan Price—a classical Greek epigrapher (among other qualifications)—was that this inscription had been done by one hand, that it likely referred to a single individual, and that it should likely be read as "Mariam also known as Mara," the presumption being that Rahmani misread an intended $\kappa\alpha\iota$ as an $\nu o\nu$ ending in the first name together with an $\eta$ $\kappa\alpha\iota$, and that the $\kappa\alpha\iota$ in this instance was intended to signify a double name. (See Figure 1.) Were that the case, it seems to me that the element (g) of our "due diligence" list above could then be supplanted by one that would now be considerably stronger still. It is worth mentioning here that the classical Greek epigrapher Roger Bagnall had earlier independently arrived at a similar reading: "Mariam, also called Mara," finding Rahmani's reading to be "not acceptable," but proposing that Mara may have been intended as a short form of Maria, although "some uncertainty remains" (quoted from a June 2007 e-mail communication). Unfortunately, the (spiraling) multiplicity of readings and interpretations for that inscription, and the nature of the relative uncertainties among them, makes it difficult to give unequivocal preference to any one of the readings, and until further work and consensus establishes at least the correct reading of this inscription (let alone any correct interpretation of it) further progress along this front seems unlikely.

One possibility that had not occurred to me was raised by a participant [Claude Matlofsky] and seems worth noting here. Namely, that the Talpiyot tomb *might* also fit the profile of the family of Jesus' brother Yoseh. Under that scenario, Yoseh would have named a son after his slain brother, but the assumption A.8 about Yoseh and Yehosef being necessarily distinct individuals would have to be suspended. It bears pointing out here that statistical 'evidence' of the nature described in my paper, even if significant, cannot automatically be used to also identify the actual persons buried in the tomb, nor any of their relationships to each other; these are separate inferential problems.

*Some opinions and concluding remarks.* During the course of this work I have had occasion to meet many of the individuals involved in this matter, including Andre Lemaire who "discovered" the James ossuary, Oded Golan who owns it and who kindly permitted me a private viewing of part of his remarkable antiquities collection, Shimon Gibson and Amos Kloner both of whom (along with the late Yosef Gath) were present at the Talpiyot find in 1980, as well as with a number of other key persons. In such meetings I tried to gather information, or at least to form impressions, about some of the nonstatistical aspects relevant to the analysis. A few such observations may be worth mentioning here. First, opinion on the authenticity of the James ossuary is divided so I have no basis in forming a judgement on that matter. Either way, if the James ossuary provenanced to the Talpiyot tomb (as some have claimed) the statistical implications would be nontrivial. However, in my opinion there is at present no credible evidence to tie that ossuary to the Talpiyot tomb. Second—although no one who has witnessed first hand the intensity that can be engendered by this subject matter would deny that such an eventuality should, at the very least, be momentarily considered— the possibility of any "cover-up" of facts by the archeologists involved strikes me as being pure fiction. The dynamics for such a thing to have taken place simply were not there. Third, a story made headlines when the widow of Yosef Gath announced that her late husband knew and had told her that he had discovered the tomb of Jesus, and that he was deeply concerned about the possible repercussions of that find. Having been present at the event during which she made this statement, I found it easy enough to gather sufficient information to lead me to be concerned that this could have been an instance of "false memory syndrome"; I am therefore inclined discount that information.

A few participants at the Princeton symposium indicated that it might be worthwhile to carry out further excavations at the Talpiyot tombsite and in particular at another immediately adjacent tomb. While it is always possible that such further work might lead to more definitive answers, it is also the case that Israeli laws are very strict about matters that pertain to disturbing burial sites. Therefore, unless evidence comes to light to invalidate the Talpiyot find, this, it seems to me, is where matters are likely to rest for some time to come.

## REFERENCES

BOVON, F. (2002). Mary Magdalene in the *Acts of Philip.* In *Which Mary?—The Marys of Early Christian Tradition* (F. Stanley Jones, ed.) 77–89. *Society of Biblical Literature Symposium Series* **19**. Society of Biblical Literature, Atlanta.

DIACONIS, P. and MOSTELLER, F. (1989). Methods for studying coincidences. *J. Amer. Statist. Assoc.* **84** 853–861. MR1134485

ILAN, T. (2002). *Lexicon of Jewish Names in Late Antiquity, Part 1*: *Palestine 330 BCE–200 CE.* Mohr Siebeck, Tubingen.